\documentstyle[12pt]{article}
\begin{document}
\baselineskip=16pt

\begin{titlepage}
\begin{flushright} ITP-SB-95-02 \\
\end{flushright}
\vspace{0.5cm}
\begin{center}{\large {\bf $sl(N)$ Onsager's Algebra \\ and Integrability } }
\\
\vspace{2cm}
D.B.Uglov \footnotemark[1] and I.T.Ivanov \footnotemark[2]    \\
\footnotetext[1]{e-mail: denis@insti.physics.sunysb.edu}
\footnotetext[2]{e-mail: iti@insti.physics.sunysb.edu}

\vspace{1cm}
Institute for Theoretical Physics, State University of New York at Stony Brook
\\ Stony Brook, NY 11794-3840, USA \\
\vspace{0.5cm}
February  , 1995
\end{center}
\vspace{2cm}
\begin{abstract}
We define an $ sl(N) $ analog of Onsager's Algebra through a finite set of
relations that  generalize  the Dolan Grady defining relations for the original
Onsager's Algebra. This infinite-dimensional Lie Algebra is shown to be
isomorphic to a fixed point subalgebra of $ sl(N) $ Loop Algebra with respect
to a certain involution. As the consequence of the generalized Dolan Grady
relations a Hamiltonian linear in the generators of $ sl(N) $ Onsager's Algebra
is shown to posses an infinite number of mutually commuting integrals of
motion.
\end{abstract}

\end{titlepage}

\def\pro #1 #2\par{\medbreak
\noindent{\bf#1 \enspace}{\sl#2}\par\medbreak}

\newtheorem{lemma}{Lemma}
\newtheorem{prop}{Proposition}
\newtheorem{cor}{Corollary}

\newcommand{\be}{\begin{eqnarray*}}
\newcommand{\ee}{\end{eqnarray*}}

\newcommand{\ben}{\begin{eqnarray}}
\newcommand{\een}{\end{eqnarray}}

\newcommand{\beqn}{\begin{equation}}
\newcommand{\eeqn}{\end{equation}}

\newcommand{\beq}{\[}
\newcommand{\eeq}{\]}

\newcommand{\ac}{\mbox{${\cal A}_N $}}
\newcommand{\act}{\mbox{$\tilde{\cal A}_N $}}
\newcommand{\St}{\mbox{$ \tilde{S} $}}
\newcommand{\e}{ {\rm {\bf e}}}

\newcommand{\w}{{\rm when}}
\newcommand{\de}{\mbox{$ \stackrel{\rm def}{=} $}}

\section{Introduction}

The seminal work of Onsager of 1944 on the exact solution of the planar Ising
Model \cite{Ons} has been a source of a considerable part of the subsequent
developments in the field of exactly solvable models in Statistical Mechanics
and Field Theory. One branch of these developments which originated with
Star-Triangle relation \cite{Ons,Wannier} and led to Yang-Baxter equation  and,
later, to theory of Quantum Groups was particularly vigorous.

Yet the Star-Triangle relation did not play any essential role in original
Onsager's solution of 2D Ising Model. Indeed it is only mentioned in
\cite{Ons}. The crucial part was played by a certain infinite-dimensional Lie
algebra which is now called Onsager's Algebra and by what one may call
associated representation theory.

This algebra  can be described by introducing the basis
$ \{ A_m , G_n \} \; , \; m = 0,\pm1,\pm2,\dots \; ; \; \\ n=1,2,\dots \;$.
Commutation relations in this basis are:
\ben
[A_l,A_m] & = & 4G_{l-m} \;\; \; l \geq m  \\
\mbox{} [G_l,A_m] & = & 2A_{m+l} - 2A_{m-l} \\
\mbox{} [ G_l,G_m ] & =  & 0
\een

Even though this algebra was at the centre of  the original Onsager's solution
of 2D Ising Model, it received substantially less attention in subsequent years
then the Star-Triangle Relation. In the context of the Ising Model the
algebraic method of Onsager was superseded by  simpler and more powerful
methods that rely on the equivalence  of the 2D Ising Model to  a
free-fermionic theory \cite{Kauf} and the dimer problem \cite{Kast}. This
caused  Onsager's Algebra to remain in a shadow for quite a long time. The
situation changed in the 1980-s when several important advances related to
Onsager's Algebra  took place. Some of these we briefly review below in order
to set up a background for the subsequent discussion.

Dolan and Grady \cite{Dolan} considered Hamiltonians $H$ of the form (in a
different notation) :
\beqn H \; = \; A_0 + k A_1 \eeqn where $k$ is a constant and $ A_0 , A_1 $ are
operators. They have shown that the following pair of conditions imposed upon
the operators $ A_0$ and $A_1$ - Dolan Grady relations -  :
\ben
\mbox{} [A_0, [A_0, [A_0 , A_1]]] & = & 16 [A_0 , A_1]  \\
\mbox{} [A_1, [A_1, [A_1 , A_0]]] & = & 16 [A_1 , A_0]
\een
are sufficient to guarantee that $ H $ belongs to an infinite family of
mutually commuting operators - integrals of motion for $H$. To be more precise,
only one of the relations (5,6) was considered in \cite{Dolan} - the second one
was produced as  a consequence of certain duality condition imposed on the
operators $A_0 , A_1 $. The Dolan Grady relations in the form (5,6) - without
the assumption of duality - were first discussed in \cite{Davies1,Davies}.
Based on the work of Dolan and Grady,  Perk \cite{Perk-Ons}  and Davies
\cite{Davies1} established moreover, that the Lie Algebra generated by two
``letters''  $ A_0 , A_1$ subject to the relations (5,6) is precisely Onsager's
Algebra as it is defined in (1-3). The  elements $ A_0 , A_1 $ of the basis $
\{ A_m , G_n \} $ are identified with $ A_0 , A_1 $ in (5,6) and all the rest
are  expressed as commutators of $ A_0 $ and $ A_1$.

The representation of the generators $ A_0 $ and $ A_1 $ which was considered
by Onsager is:
\ben A_0 \; = \; \sum_{i=1}^L \sigma_i^x  &,  &  A_1 \; = \; \sum_{i=1}^L
\sigma_i^z \sigma_{i+1}^z \;. \een
and after substitution in (4) gives the  Hamiltonian of Transverse Ising Chain.
This Hamiltonian is defined on a periodic chain of length $L$ ; and $
\sigma_i^{x,z} $ are Pauli matrices representing local spins on a site with a
number $i$. It is well known that Jordan-Wigner transformation brings this
Hamiltonian to a free-fermionic form \cite{Kauf}.

An important question -  whether there are other representations of the
relations (5,6) leading to nearest-neighbor spin Hamiltonians that  are not
free  - was answered affirmatively by von Gehlen and Rittenberg in 1985
\cite{Rittenberg} who found a family of such representations. For every integer
$ M \geq 2 $ they define:
\ben A_0 \;  = \; \frac{4}{M} \sum_{i=1}^L \sum_{n=1}^{M-1}\frac{X_i^n}{1-
\omega^{-n}} & , &
A_1 \;  = \; \frac{4}{M} \sum_{i=1}^L \sum_{n=1}^{M-1}\frac{Z_i^n
Z_{i+1}^{M-n}}{1- \omega^{-n}}
\een where $ X_i , Z_i $ are local $Z_M$ spin operators satisfying: $
[X_i,X_j]=[Z_i,Z_j]=0\;,\; Z_iX_j = \omega^{\delta_{ij}}X_jZ_i \;,\; Z_i^M =
X_i^M = I $; and $ \omega = {\rm exp}(2\pi i/M) $. When $M=2$ this
representation coincides with  Onsager's original representation (7) for Ising
Model. For arbitrary integer $ M  $ the spin-chain Hamiltonians of the form (1)
with $A_0 , A_1 $ given by (8)  were later shown to be certain - so called {\it
Superintegrable } \cite{Albertini,Baxter2,BBP} - specializations of the
spin-chains generated by 2D  Chiral Potts Model [10-14]. von Gehlen and
Rittenberg also observed numerically certain Ising-like structure of some
eigenvalues in the spectrum of these Hamiltonians. In \cite{Albertini,CPSpec}
the $ M=3 $ case was solved analytically and this structure was shown to hold
for all eigenvalues. This Ising-like form of eigenvalues was later rigorously
established by Davies \cite{Davies} to be a consequence of Onsagers's Algebra;
Davies proved that all eigenvalues of a
 Hamiltonian of the form (4) defined in a finite-dimensional Hilbert space fall
into multiplets parameterized by: two real numbers $ \alpha , \beta $ ,
positive integer $ n $ , $n$ real numbers $\theta_i$ and $n$ nonnegative
integers $s_i$; eigenvalues which  belong to such a multiplet being given by
the formula:
\beqn
 \alpha + k \beta + \sum_{j=1}^n 4m_j \sqrt{1 + k^2 + 2k cos\theta_j}
\;\;\;,\;\; m_j = -s_j, -s_j + 1, \dots , s_j .
\eeqn

Classification of the finite-dimensional representations of Onsager's Algebra
which leads to this form of the eigenvalues was carried out by Davies
\cite{Davies} and, subsequently, by Roan \cite{Roan}.

Onsager's Algebra by itself does not define the parameters $ \{
\alpha,\beta,\theta_i,s_i \} $ entering into the eigenvalue formula (9). To
find these for a given representation of the generators $ A_0 , A_1 $ in some
Hilbert Space ${\cal H}$, one needs to find the decomposition of ${\cal H}$
into irreducible subrepresentations of Onsager's Algebra \cite{Davies}.

For the Superintegrable Chiral Potts Hamiltonians given by (8) the complete
spectrum of eigenvalues has been found in \cite{CPSpec} for the 3-state case $
M=3 $ with the aid of a certain cubic relation satisfied by the Transfer-Matrix
of Chiral Potts Model. For general $ M $ the eigenvalues of the ground-state
sector were obtained in \cite{Baxter2} by use of an inversion identity for the
same Transfer-Matrix , in \cite{Tarasov} this result was extended to other
eigenvalues. Further results related to the superintegrable Chiral Potts Model
can be found in \cite{SCP}.

Now we come to the motivation and the subject of the present paper. Since the
work of Onsager it was  known that there exists an intimate relationship
between Onsager's Algebra and $ sl(2) $. This relationship was  clarified by
Roan \cite{Roan} who  built on the earlier work of Davies
\cite{Davies1,Davies}. Namely, Roan has shown that the Onsager's Algebra given
by (5,6) ( or equivalently by (1-3) ) is isomorphic to the fixed-point
subalgebra of $ sl(2) $-Loop   Algebra $ {\cal L}(sl(2)) $ ( or ,
alternatively, of its central extension $ A_1^{(1)} $ \cite{Kac}) with respect
to the action of a certain involution. Indeed one can easily guess that there
should be a connection between Onsager's Algebra and Kac-Moody Algebra $
A_1^{(1)} $ looking at the Dolan Grady relations whose left-hand sides coincide
with the left-hand sides of the Serre relations for $ A_1^{(1)} $ \cite{Kac}.

 This obviously raises the question: whether one can find a generalization of
Onsager's Algebra related in some way to $ {\cal L}(sl(N)) $ for $ N\geq 3$ and
to other Loop or Kac-Moody Algebras ? In particular: can one find some
generalizations of the Dolan Grady conditions (5,6) leading to integrability in
the sense of existence of infinite series of integrals of motion in involution?

The aim of this paper is to propose such a generalization related to
$sl(N)$-Loop Algebra for $N \geq 3$. We reserve the discussion of the
generalizations related to Loop Algebras over other classical Lie Algebras for
a future publication.

   Let us summarize the results. We consider Hamiltonians of the form:   \beqn
 H \; = \; k_1 \e_1 + k_2 \e_2 + \dots + k_N \e_N  \;\;\; , \;\;\;\; N \geq 3
\eeqn where $k_i$ are some arbitrary numerical constants and $ \e_i $ are
linear operators.

We find that if the operators $ \e_i $ satisfy certain commutation relations (
generalized Dolan Grady relations given by the formulas (11-12) below ) then
the Hamiltonian $ H $ is a member of an infinite family of mutually commuting
integrals of motion (see formula (48)). Each of these integrals is linear in
the coupling constants $ k_i $ and is explicitly expressed  in terms of $
\e_i$. This property  follows from the fact that the Lie Algebra generated by $
\e_i $ subject to the relations (11,12) ( we call it $ sl(N) $ Onsager's
Algebra and denote it by \ac\ ) is isomorphic to the fixed-point subalgebra of
$ sl(N) $ Loop Algebra under the action of a certain involution.

The important problem is to find examples of interesting Hamiltonians of the
form (10).
The only example of a  Hamiltonian satisfying the generalized Dolan Grady
relations that we have been able to find so far is the Hamiltonian of an
inhomogeneous periodic Ising chain of length $L$. To write down such a
Hamiltonian we first define a sequence of operators:
\beq
g_{2k-1}\; = \; \frac{1}{2}\sigma^x_k \; \; , \;\;g_{2k}\; = \;
\frac{1}{2}\sigma^z_k \sigma^z_{k+1} \;\;, \;\;\; k = 1,2,\dots,L
\eeq
Then we take the operators $ \e_i $ entering into the Hamiltonian to be given
by: $ \e_i \:=\: g_i $ for $ i=1,2,\dots,2L $. These operators satisfy (11,12)
for the $ sl(2L) $ Onsager's Algebra. After substitution of these operators
into (10) we get the completely inhomogeneous Transverse Ising Chain with
inhomogeneities $ k_i $. If we take $ 2L = mN $ for some integer $m$, then
defining the operators $ \e_i $ by: $ \e_i \: =  \: \sum_{s=0}^{m-1} g_{Ns+i} $
for $ i = 1,2,\dots, N $ we get a representation of $ sl(N) $ Onsager's
Algebra. The Hamiltonian corresponding to this representation is the
Hamiltonian of the Transverse Ising Chain with periodic inhomogeneities. These
Hamiltonians have been  known for a long time to be  free-fermionic
\cite{Book,Perk}. So the outstanding unsolved problem is to find
representations of $ sl(N) $ Onsager Algebra that give rise to models that
cannot be mapped onto free-fermions  , or to prove that such representations do
not exist.

Now we outline the contents of the paper. In sec. 2 we give the definition of
the main object of our study: $sl(N)$ Onsager's Algebra which we denote by \ac\
. We specify the Algebra by giving $ N $ generators and a finite number of
defining relations that generalize the Dolan Grady conditions.
 In sec. 3 we discuss an involution of the $sl(N)$ Loop Algebra and its
fixed-point subalgebra \act\ , as we shall see later this subalgebra is
isomorphic to \ac\ . In order to prepare the proof of this isomorphism in
subsequent sections we introduce two convenient bases in \act\  .
In sec. 4 we study the structure of the algebra \ac\ and prove a Proposition
which gives a set of elements that span \ac\ as a linear space. We achieve this
result by computing commutators of the generators with the elements of this
set.
In sec. 5 the isomorphism of the Lie Algebras \ac\ and \act\ is established.
Due to the results established in sec.3 this gives us a basis of the Algebra
\ac\ and commutation relations among the elements of this basis.
In sec. 6 the knowledge of the basis and commutation relations enables us to
find an infinite family of mutually commuting elements in the Lie Algebra \ac\.
The Hamiltonian (10) is one of these integrals of motion.
%Sec. 7 is devoted to a conclusion.

\section{Definition of \ac\ - $sl(N)$ analog of Onsager's Algebra }

In this section we introduce the Lie algebra \ac\ which is a generalization of
the original Onsager's algebra to the $sl(N)$, $ N \geq 3 $ case. The relation
of this algebra to $sl(N)$ or, more precisely, to the loop algebra of Laurent
polynomials with values in $sl(N)$ will be explained later in sec. 5.
 In order to define the algebra \ac\ it is convenient to consider the Dynkin
graph of the type $A_{N-1}^{(1)}$:
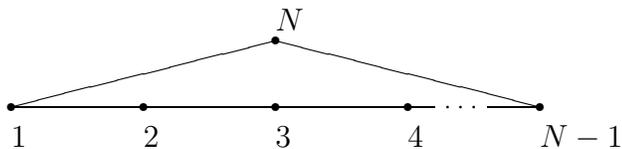
\begin{center}
\begin{picture}(205,60)(5,-20)
\put(0,0){\line(4,1){100}}
\put(100,25){\line(4,-1){100}}
\put(0,0){\line(1,0){160}}
\put(180,0){\line(1,0){20}}
\put(100,25){\circle*{3.0}}
\multiput(0,0)(50,0){4}{\circle*{3.0}}
\put(200,0){\circle*{3.0}}
\multiput(165,0)(5,0){3}{\circle*{0.5}}
\put(100,29){$N$}
\put(0,-15){$1$} \put(50,-15){2} \put(100,-15){3} \put(150,-15){4}
\put(200,-15){$N-1$}
\end{picture}
\end{center}
We label the vertices of this graph by $i$ ranging from 1 to $N$ , $ i+N \de\ i
$. To a vertex with a label $i$ we attach a letter $\e_i$ . Then we define \ac\
to be a complex Lie algebra generated by the letters $ \e_i\;,\; i=1,2,\dots,N
$ subject to the following defining relations:
\ben
[\e_i,[\e_i,\e_j]] & = & \e_j \; \; \; \;\mbox{ if $i$ and $j$ are adjacent
vertices} \\
\mbox{} [\e_i,\e_j] & = & 0 \;\;\;\;\mbox{ if $i$ and $j$ are not adjacent }
\een
As a linear space the algebra \ac\  is a linear span of all multiple
commutators of $\e_i$ between themselves  taken modulo the relations (11,12).

Let us now introduce some notations. We shall often be working with multiple
commutators nested to the right, that is expressions of the form: $ [ a_{k_1},
[ a_{k_2}, [a_{k_3}, \dots [ a_{k_{m-1}}, a_{k_m}] \dots ]]] $ where $ a_{k_i}
$ are some elements in the Lie Algebra \ac\ . For such a commutator we shall
use the notation:
\beq [ a_{k_1}, [ a_{k_2}, [a_{k_3}, \dots [ a_{k_{m-1}}, a_{k_m}] \dots ]]]\;
\de\ \;[ a_{k_1},  a_{k_2}, a_{k_3}, \dots , a_{k_{m-1}}, a_{k_m} ] \eeq
In multiple commutators in which the generators $ \e_i $ appear, we shall
replace the symbol $ \e_i $ by $ i $, for example:
\beq
    [\e_5,[\e_1,[\e_2,\e_5]]]\; \de \; [5,1,2,5] \; \de \; [5,[1,2,5]] \; \de
\; {\rm etc.} \eeq
% Among multiple commutators of $\e_i$ we will find it useful to consider those
%%ones nested to the right, i.e. expressions of the form%: $
%%[\e_{i_1},[\e_{i_2},[ \dots [\e_{i_{m-1}},\e_{i_m}]\dots ]]] $. For such a
%%commutator we will use notation:
%\beq [\e_{i_1},[\e_{i_2},[ \dots [\e_{i_{m-1}},\e_{i_m}]\dots ]]]\; \de\
%%\;[i_1,i_2,\dots,i_m]
%\eeq
Now we define elements of \ac\ which will play an important role in subsequent
discussion. These elements are denoted by $S_k(r)$ and defined as follows:
\ben
S_k(r) & \de\ & [k,k+1,k+2,\dots,k+r-1] \;\;\;  k\;=\;1,2,\dots,N \;;\;
r\;=\;1,2,3,\dots
\een
We shall call such an element {\it a string }  of  length  $r$. Strings are
cyclic in their sub-indices: $ S_{k+N}(r) \; = \; S_k(r)$.  Strings of length 1
are the generators of \ac\ : $ S_k(1) \; = \; \e_k $. String of length 0 is by
convention equal to zero. As we will see in sec. 4, the whole algebra \ac\ is
spanned by strings as a linear space. Strings are linearly independent except
that the sum of all closed strings (i.e. strings whose length is divisible by
$N$ ) of a given length vanishes. This will be established in sec. 5.

The algebra \ac\ has an automorphism of order $N$ which we will use later. This
automorphism which we denote $C$ is defined by cyclic permutation of the
generators:
\ben
C:\; \e_i & \rightarrow & \e_{i+1}
\een
The automorphism $C$ is quite useful in computations of commutators, since the
action of this automorphism on strings is again  cyclic permutation:
\ben
C: \; S_i(r) & \rightarrow & S_{i+1}(r)
\een

The obvious questions one can ask about the Lie Algebra \ac\ are: what is a
basis of this algebra and what are commutation relations among elements of this
basis. These questions are answered in the sections 4 an 5. There we shall
establish the isomorphism between \ac\ and the Lie Algebra \act\ which we
define and describe in the next section.

%%%%%%%%%%%%%%%%%%%%%%%%%%%% ALGEBRA A~
\section{The Loop Algebra ${\cal L}(sl(N))$, its involution and the fixed-point
subalgebra \act\ .}

  As we shall see later, the Lie algebra \ac\ introduced in the previous
section is closely related to ${\cal L}(sl(N))$ - the $sl(N)$ loop algebra. In
this section we describe a certain involution $\omega$ of ${\cal L}(sl(N))$ and
the Lie subalgebra \act\ of ${\cal L}(sl(N))$ on which the action of $\omega$
is reduced to the identity (``fixed-point subalgebra `` of $\omega$). In
subsequent sections we shall prove that the algebras \ac\ and \act\ are
isomorphic and shall describe this isomorphism.

The $sl(N)$ loop algebra: ${\cal L}(sl(N)) \; = \; {\rm {\bf
C}}[t,t^{-1}]\otimes sl(N)$ has the basis $ \{ E_{ij}^{(n)} \;, \; H_k^{(n)} \}
\; \; 1\leq i \neq j \leq N \: ; \:1\leq k \leq N-1 \;\;;\;\; n = 0,\pm 1 , \pm
2, \dots \;$. In the $ N\times N $-matrix realization of $sl(N)$ the elements
of this basis have the explicit form:
\be  E_{ij}^{(n)} \;=\; t^n E_{ij}\:, &   H_k^{(n)}\;=\; t^n H_k \;=\; t^n (
E_{kk}-E_{k+1k+1} ) &
\ee where $ E_{kl} $ is $ N\times N $ matrix whose all entries are zero except
entry $(kl)$ which is equal to 1. The loop algebra has a linear involutive
automorphism $ \omega $ , $ \omega^2 = id $, given by:
\ben
\omega : E_{ij}^{(n)} & \rightarrow & (-1)^{i+j+1+nN}E_{ji}^{(-n)} \\
\omega : H_i^{(n)} & \rightarrow & (-1)^{1+nN}H_i^{(-n)}
\een
This involutive automorphism $\omega$ is a product of two involutions:
$\omega_1$ and $\omega_2$.
The first of these is an involution of the algebra of Laurent polynomials:
\beq
\omega_1 : \; t^n \; \rightarrow \; (-1)^{nN}t^{-n}
\eeq
The second one is an involution of $sl(N)$:
\be
\omega_2 : E_{ij} & \rightarrow & (-1)^{i+j+1}E_{ji} \\
\omega_2 : H_i & \rightarrow & -H_i
\ee

It is easy to convince oneself that the subspace $ \act\ \; \in \;  {\cal
L}(sl(N)) $ on which the involution $\omega $ acts as identity operator, is a
Lie subalgebra.
We can easily find the basis of this fixed-point subalgebra \act\ ; it is
formed by vectors $ \{ A_{ij}^{(n)} , G_i^{(n)}\}$ :
\ben
A_{ij}^{(n)} & = & E_{ij}^{(n)} + (-1)^{i+j+1+nN}E_{ji}^{(-n)} \; \; \; \;
1\leq i<j \leq N \;,\; n=0,\pm 1,\pm 2,\dots  \\
G_i^{(n)} & = & H_i^{(n)} + (-1)^{nN+1}H_i^{(-n)} \; \; \;\; 1\leq i \leq N-1
\;,\, n=1,2,\dots
\een


The commutation relations of \act\ in the basis $ \{ A_{ij}^{(n)}\;,\;
G_i^{(n)} \} $ follow immediately from the commutation relations of ${\cal
L}(sl(N))$:
\ben
[A_{ij}^{(m)},A_{kl}^{(n)}] & = & \delta_{jk}A_{il}^{(m+n)} -
\delta_{il}A_{kj}^{(m+n)} + \nonumber \\
& & \delta_{ik}(-1)^{i+j+1+mN}\theta (j<l)A_{jl}^{(n-m)} +
\delta_{ik}(-1)^{i+l+nN}\theta (l<j)A_{lj}^{(m-n)} + \nonumber \\
& & \delta_{jl}(-1)^{k+l+1+nN}\theta (i<k)A_{ik}^{(m-n)} +
\delta_{jl}(-1)^{i+l+mN}\theta (k<i)A_{ki}^{(n-m)} + \nonumber \\
& & \delta_{ik} \delta{jl} (-1)^{i+j+1+nN}\sum_{s=i}^{j-1} G_s^{(m-n)} \; \; \;
\; m\geq n      \\ \mbox{}
[G_i^{(m)},A_{kl}^{(n)}] & = & ( \delta_{ik}- \delta_{ki+1} - \delta_{li} +
\delta_{li+1})(A_{kl}^{(m+n)}-(-1)^{mN}A_{kl}^{(n-m)}) \\ \mbox{}
 [G_i^{(m)}, G_j^{(n)}] & = & 0
\een
here $\delta_{ik}$ is the Kronecker symbol and $\theta (x)$ is the following
function: $ \theta (x) \; = \; 1(0) \; $ if $x$ is true(false).


We will also need another basis in the algebra \act. We shall denote the
elements of this new basis by
symbols $\St_i(r)$ where $ r=1,2,\dots \; $; and $ 1 \leq i \leq N $ if $r$ is
not divisible by $N$ ; and $ 1 \leq i \leq N-1 $ if $ r=Nm $ for some positive
integer $m$. The explicit expressions for the elements $\St_i(r)$ are as
follows:
\ben
\St_i(k) & = & \left \{ \begin{array}{ll}
                        E_{ii+k} + (-1)^{k+1}E_{i+ki} & \mbox{ if $ i+k \leq N
$} \\
                        t E_{ii+k-N} + (-1)^{k+1}t^{-1} E_{i+k-Ni} & \mbox{ if
$ i+k \geq N+1$}
                        \end{array}   \right.
\een for $ 1\leq k\leq N-1 $;

\ben
\St_i(Nm+k) & = & \left \{ \begin{array}{r}
    (t-(-1)^Nt^{-1})(t+(-1)^Nt^{-1})^{m-1}(E_{ii+k} + (-1)^kE_{i+ki}) \\ \mbox{
if $ i+k \leq N $} \\
 (t-(-1)^Nt^{-1})(t+(-1)^Nt^{-1})^{m-1}(t E_{ii+k-N} + (-1)^{k}t^{-1}
E_{i+k-Ni}) \\ \mbox{ if $ i+k \geq N+1$}
                        \end{array}   \right.
\een for $ 1\leq k\leq N-1 \; , \; m \geq 1 $;

\ben
\St_i(Nm) & = & (t-(-1)^Nt^{-1})(t+(-1)^Nt^{-1})^{m-1}(E_{ii}-E_{i+1i+1})
\een for $ m \geq 1 \; , \; 1 \leq i \leq N $.
Note that the elements defined by the last formula are linearly dependent:  $
\St_1(Nm)+\St_2(Nm)+\dots+\St_N(Nm)\;=\;0 $.


We can express the elements of the basis $\{ A_{ij}^{(n)}\, , G_i^{(n)} \}$ in
terms of the basis $\{ \St_i(r) \}$ with the aid of the recursion relations:
\be
A_{ij}^{(0)} & = & \St_i(j-i) \\
A_{ij}^{(-1)} & = & (-1)^{1+N+i+j} \St_j(N+j-i) \\
A_{ij}^{(m)} & = & (-1)^N A_{ij}^{(m-2)} + \Phi_i(Nm,j-i) -
(-1)^{i+j}\Phi_j(N(m-1),N+i-j) \\
A_{ij}^{(-m-1)} & = & (-1)^N A_{ij}^{(-m+1)} +(-1)^{mN}\Phi_i(N(m-1),j-i) \\
 & & - (-1)^{(m+1)N+i+j}\Phi_j(Nm,N+i-j) \\
G_i^{(m)} & = & \Phi_i(Nm,0)
\ee for $ m \geq 1 $. The vectors $ \Phi_i(Nm,s) $ are given in terms of $
\St_i(r) $ by the formula:
\be
\Phi_i(Nm,s) & = & \sum_{r=1}^m c(m,r) \St_i(Nr+s) \; \; \; \; 0\leq s\leq N-1
\ee where coefficients $ c(m,r) $ are defined by the recursion relation:
\be
c(m+1,r) & = & (1-\delta_{1r})c(m,r-1) - (1-\delta_{mr})(1-\delta_{m+1r})(-1)^N
c(m-1,r) \\
c(1,1) & = & 1
\ee

The basis of vectors $ \{ \St_i(r) \} $ which we described in this section will
be used in the proof of an isomorphism of the algebras \ac\ and \act\ .

\section{Structure of the algebra \ac\ }

In this section we study the structure of the algebra \ac\ in some detail. The
main result that we establish is formulated as the following proposition:
\begin{prop}
The  Lie algebra \ac\ is spanned by the set of  strings $\{ S_i(r) \}\; 1\leq
i\leq N\;,r=1,2,\dots $ as a linear space.
\end{prop}
Notice that it is not true that all strings are linearly independent.

In order to prove this proposition we shall first compute commutators of the
generators of \ac\ with all strings, that is  the commutators of the form:
\be
[\e_i,S_j(r)] & \de\ & [i,S_j(r)]
\ee where $ 1 \leq i,j \leq N $ and $ r=1,2,\dots $.
Due to the existence of the cyclic automorphism $C$ it is sufficient to compute
$ [1,S_j(r)] $ for all $ j $ and $r$ ; the rest of the commutators $ [i,S_j(r)]
$ is then immediately obtained by application of $C$.
The result which we get computing $ [1,S_j(r)] $ is summarized in the Lemma .
The distinctive feature of the strings which emerges from the result of the
Lemma  is that a commutator of a generator with a string is again a string.

%\begin{lemma}
\pro Lemma
The following relations hold in \ac\ for $m \geq 0$:
\[   [1,S_k(Nm+r)] = \]

{\rm 1. If} $ 1\leq k \leq N $ , $ 1\leq r \leq N-1 $  {\rm and}   $ k+r \leq N
$,
\begin{eqnarray*}
{\rm a}_m).\hspace{0.5cm}  &  -2S_1(Nm) &  \w\ \;   k=1, r=1  \\
{\rm b}_m).\hspace{0.5cm}  &   S_2(Nm + r-1) & \w\ \;   k=1, r \geq 2 \\
{\rm c}_m).\hspace{0.5cm}  &   S_1(Nm + r+1) &  \w\ \;  k=2 \\
{\rm d}_m).\hspace{0.5cm}  &    0  &  \w\ \;  k \geq 3 \\
\end{eqnarray*}

{\rm 2. If} $ 1\leq k \leq N $ , $ 1\leq r \leq N-1 $  {\rm and}   $ k+r \geq
N+1 $,
\begin{eqnarray*}
{\rm e}_m).\hspace{0.5cm}  &  -S_k(Nm + r+1) &  \w\ \;   k+r = N+1, k \neq 2
\\
{\rm f}_m).\hspace{0.5cm}  &   S_1(Nm + r+1) & \w\ \;   k+r = N+1, k = 2 \\
{\rm g}_m).\hspace{0.5cm}  &   -S_k(Nm + r-1) &  \w\ \;  k+r=N+2 \\
{\rm h}_m).\hspace{0.5cm}  &    0  &  \w\ \;  k+r \geq N+3 \\
\end{eqnarray*}

{\rm 3. If} $ 1\leq k \leq N $ {\rm and}   $ r = N $,
\begin{eqnarray*}
{\rm i}_m).\hspace{0.5cm}  &  -2S_1(Nm + N+1) &  \w\ \;   k = 1   \\
{\rm j}_m).\hspace{0.5cm}  &   S_1(Nm + N+1) & \w\ \;    k = 2 \\
{\rm k}_m).\hspace{0.5cm}  &   S_1(Nm + N+1) &  \w\ \;  k = N \\
{\rm l}_m).\hspace{0.5cm}  &    0  &  \w\ \; 3 \leq  k \leq N-1 \\
\end{eqnarray*}

%\end{lemma}

{\it Proof:}

We shall prove the Lemma using induction in $m$. First we establish the base of
the induction by proving relations $ {\rm a}_0 $ through $ {\rm l}_0 $, and
then show, that relations $ {\rm a}_m $  through $ {\rm l}_m $ entail  $ {\rm
a}_{m+1}$ through $ {\rm l}_{m+1} $. At each elementary step we employ either
the defining relations of \ac\ or the Jacobi identity or skew-symmetry of the
commutator. The proof given below is valid for $N \geq 4$. Proof for $N=3$
differs in some details and is omitted here.

1.) Proof of the induction base. We compute the commutator $ [1,S_k(r)] $ when
$ 1 \leq k \leq N $ and $ 1 \leq r \leq N $.

$\underline{ Case \; {\rm a_0} }$ :  $ k=1 , r=1 $.

$[1,S_1(1)] \;  \de\ \; [1,1] \; = \; 0  \;  \de\ \;   -2 S_1(0) $.
\vspace{0.4cm}

$\underline{ Case \; {\rm d_0} }$ :  $ k \geq 3 , k+r \leq N , r \leq N-1$.

$x \; \de\ \; [1, S_k(r)]\; \de\ \;[1,k,k+1,\dots,k+r-1].$  Since $ k \geq 3 $
 and $  k+r \leq N-1 $, 1 commutes with all $k,k+1,\dots, k+r-1$. Hence $ x=0
$.
\vspace{0.4cm}

$\underline{ Case \; {\rm b_0} }$  :  $ k = 1 ,\; r \geq 2,\; k+r \leq N $.

$x \; \de\ \; [1, S_k(r)]\; \de\ \;[1,1,2,\dots,r].$  Since $ r \leq N-1 $, $
x\; =\; [ 1,[1,2],3,\dots,r] \;=\; [2,3,\dots,r] \;  \de\ \; S_2(r-1) $ .
\vspace{0.4cm}

$\underline{ Case \; {\rm h_0} }$  :  $ k+r \geq N+3,\; r \leq N-1 $.

$x\; \de\ \; [1,S_k(r)]\; \de \; [1,k,k+1,\dots,N,1,2,\dots,k+r-1-N] .$ Since $
2 \leq k+r-1-N \leq N-2,$
\beq
x\; = \; [k,k+1,\dots,N-1,[1,N],1,2,\dots,k+r-1-N] +
[k,k+1,\dots,N,1,1,2,\dots,k+r-1-N].
\eeq
Denote the first (second) summand in the right-hand side of the above formula
by $a$ ($b$). Then we find that:
\be
a & = & -[k,k+1,\dots,N,2,\dots,k+r-1-N] + \\ & &
[k,k+1,\dots,N-1,1,[1,N],2,\dots,k+r-1-N] \\
& = &  \; -[k,k+1,\dots,N,2,\dots,k+r-1-N] + \\ & &
[k,k+1,\dots,N-1,1,1,N,2,\dots,k+r-1-N] \\
& &   - [k,k+1,\dots,N-1,1,N,1,2,\dots,k+r-1-N].
\ee
The first two summands above and the commutator $b$ vanish since $N$ commutes
with all elements standing on its right in these expressions. Hence $ a \; = \;
-x $, because the leftmost element 1 in the third summand above commutes with
all elements standing on its left. Therefore $ x = a + b = -x $ and $ x = 0 $.
\vspace{0.4cm}

$\underline{ Case \; {\rm e_0} }$  :  $ k+r = N+1,\; k \geq 3 ,\; r \leq N-1 $.
\be
[1,S_k(r)] & \de\ &  [1,k,k+1,\dots,N-1,N]\;= \\
-[k,k+1,\dots,N-1,N,1] & \de\ & -S_k(r+1) .
\ee

$\underline{ Case \; {\rm g_0} }$  :  $ k+r = N+2,\;  r \leq N-1 $.
\be
[1,S_k(r)] & \de\ &  [1,k,k+1,\dots,N-1,N,1]\;= \\
-[k,k+1,\dots,N-1,1,1,N] & = & -[k,k+1,\dots,N-1,N]\; \de \;  -S_k(r-1) .
\ee

$\underline{ Case \; {\rm i_0} }$  :  $ k=1\;, r = N $.
\be
x &  \de\ & [1,S_1(N)]\; \de \; [1,1,2,\dots,N] \\
  &  = & [1,[1,2],3,\dots,N] - [1,2,\dots,N,1].
\ee
Denote the first(second) summand in the right-hand side of the above formula by
$a$ ($-b$). Then :
\be
a &= &[2,3,\dots,N] + [[1,2],1,3,\dots,N] \\
&=& [2,3,\dots,N] - [1,2,3,\dots,N,1]+ [2,1,3,\dots,N,1] \\
 &= & - [1,2,3,\dots,N,1] \; = \; -b
\ee
Hence: $ x\; =\; a - b\; = \; -2[1,2,3,\dots,N,1]\; \de\ \; -2S_1(N+1) $.
\vspace{0.4cm}

$\underline{ Case \; {\rm k_0} }$ : $ k=N\;, r=N $.
$x\; \de\ \; [1,S_N(N)]\; \de\ [1,N,1,S_2(N-2)]\; = \; a + b $. Where $ a\;=\;
[[1,N],1,S_2(N-2)] $ and $ b\;=\; [N,1,S_1(N-1)]$. Using the defining relations
of \ac\ and the Jacobi identity we find that:
\beq
a\;=\;-[N,S_2(N-2)]+[1,1,N,S_2(N-2)]-[1,N,1,S_2(N-2)].
\eeq
Using the already proven relation $ {\rm e_0} $ and the automorphism $C$ one
finds that $ [N,S_2(N-2)]\;=\;-S_2(N-1)$. Hence $ a\;=\; S_2(N-1) - [1,S_1(N)]
- [1,S_N(N)]. $ The relation ${\rm i_0}$ then leads to $ a\;= S_2(N-1) +
2S_1(N+1) - x $. The already proven relation ${\rm b_0}$ gives: $b\;=\;
[N,S_2(N-2)]\;=\;-S_2(N-1)$. Hence we obtain $ x\; =\; a+b \;=\; 2S_1(N+1)-x $,
and $ x\;=\; S_1(N+1)$.
\vspace{0.4cm}

$\underline{ Case \; {\rm l_0} }$ : $ 3 \leq k \leq N-1 \;, r=N $.

$[1,S_k(N)]\;=\;[1,k,S_{k+1}(N-1)]\;=\;[k,1,S_{k+1}(N-1)]\;=\;0$ , since the
internal commutator in the last formula vanishes due to the already proven
relation $ {\rm h_0} $.

The rest of the cases, i.e. $ {\rm c_0}, {\rm f_0} $ and ${\rm j_0}$ are
immediate by the definition of the elements $S_i(r)$.

The induction base is proven.

2). Now we prove the induction step: the relations ${\rm a_{m+1}},\dots, {\rm
l_{m+1}}$ follow from the relations    ${\rm a_{m}},\dots, {\rm l_{m}}$.

%%%%%%%%%%%%%%%%%% THE FIRST SPECIAL CASE
\vspace{0.4cm}

$\underline{ Case \; {\rm a_{m+1}} }$ :  $ k=1 , r=1 $.

$x\;\de\ \; [1,S_1(N(m+1)+1)]\;\de\ \; [1,1,2,S_3(Nm+N-1)]\;=\;
[1,[1,2],S_3(Nm+N-1)]+[1,2,1,S_3(Nm+N-1)].$ Denote the first(second) summand in
the right-hand side of the last formula by $ a (b) $. Then using the defining
relations and the Jacobi identity we obtain:
\beq
a\;=\; [2,S_3(Nm+N-1)]+[[1,2],1,S_3(Nm+N-1)].
\eeq
Denoting the second summand in the last formula by $a_2$ and using the identity
$ {\rm g_m} $ to compute the commutator $ [1,S_3(Nm+N-1)] $ , we arrive at the
equation:
\beq
a_2 \;= \; -[1,2,S_3(Nm+N-2)]+[2,1,S_3(Nm_N-2)].
\eeq
Now we apply the identity $ {\rm e_m} $ together with the definition of a
string and find that: $a_2 \; = \; -S_1(Nm+N) - [2,S_3(Nm+N-1)]$.

Using ${\rm g_m}$ one gets $ b\;=\; -S_1(Nm+N)$. Putting expressions for $a$
and $b$ together we find:
\beq
x\;=\;a+b\;=\;-2S_1(N(m+1)).
\eeq

%%%%%%%%%%%%%%%%%%%%%%%% THE FIRST GAP

$\underline{ Case \; {\rm d_{m+1}} }$ :  $ k \geq 3 , k+r \leq N , r \leq N-1$.

$x \; \de\ \; [1,S_k(N(m+1)+r)] \; \de\ \;
[1,k,k+1,\dots,N,1,2,\dots,k-1,S_k(Nm+r)].$ The relation ${\rm d_m}$ gives: $x
\; = \;
[k,k+1,\dots,N-1,[1,N],1,2,\dots,k-1,S_k(Nm+r)]+[k,k+1,\dots,N,1,[1,2],3,\dots,k-1,S_k(Nm+r)]$. Denote the first(second) summand in $x$ by $a(b)$. Then the defining relations, the Jacobi identity and ${\rm d_m}$ give: $b\;=\; [k,k+1,\dots,N,2,3,\dots,k-1,S_k(Nm+r)]$; and $ a\;=\; -b + [k,k+1,\dots,N-1,1,[1,N],2,3,\dots,k-1,S_k(Nm+r)]$. For $x$ then we obtain:
\be
x \;= &  a+b\;= & [k,k+1,\dots,N-1,1,1,N,2,3,\dots,k-1,S_k(Nm+r)] \\
      & & -[k,k+1,\dots,N-1,1,N,1,2,\dots,k-1,S_k(Nm+r)].
\ee
Since in the second summand above the leftmost 1 commutes with all elements
standing on its left we can carry this generator 1 to the very left. In the
first summand above $N$ commutes with all elements standing on its right up to
$S_k(Nm+r)$. Hence we obtain:
\beq
2x\;=\; [k,k+1,\dots,N-1,1,1,2,3,\dots,k-1,N,S_k(Nm+r)].
\eeq
Let us now compute the commutator $ [N,S_k(Nm+r)] $ standing to the very right
in the above expression. The identities ${\rm d_m}\;,\;{\rm e_m}$ together with
the application of the automorphism $C$ give: $ [N,S_k(Nm+r)]\;
=\;-\delta_{r+k,N}S_k(Nm+r+1) $. Therefore
\be
2x &= & -\delta_{r+k,N}[k,k+1,\dots,N-1,1,1,2,\dots,k-1,S_k(Nm+r+1)] \\
   &= & -\delta_{r+k,N}[k,k+1,\dots,N-1,1,S_1(Nm+r+k)] \\
   &= & -\delta_{r+k,N}[k,k+1,\dots,N-2,1,N-1,S_1(Nm+N)]
\ee
The relation ${\rm l_m}$ applied (together with the automorphism $C$) to the
commutator $[N-1,S_1(Nm+N)]\;$ gives $ x\;=\;0. $

%%%%%%%%%%%%%%%%%%%%%%%%%%%%% THE THIRD SPECIAL CASE
\vspace{0.4cm}

$\underline{ Case \; {\rm b_{m+1}} }$  :  $ k = 1 ,\; r \geq 2,\; k+r \leq N $.

$
[1,S_1(N(m+1)+r)]\;=\;[1,1,2,S_3(N(m+1)+r-2))]\;=\;[1,[1,2],S_3(N(m+1)+r-2)]+[1,2,1,S_3(N(m+1)+r-2)].$
The commutator $[1,S_3(N(m+1)+r-2)]$ is equal to zero either because of the
already proven relation ${\rm d_{m+1}}$ or, when $r=2$, because of the relation
${\rm l_m}$. Consequently $ [1,S_1(N(m+1)+r)]\;=\;
[2,S_3(N(m+1)+r-2)]\;\de\ \; S_2(N(m+1)+r+1).$

%%%%%%%%%%%%%%%%%%%%%%%%%%%% THE SECOND GAP
\vspace{0.4cm}

$\underline{ Case \; {\rm h_{m+1}} }$  :  $ k+r \geq N+3,\; r \leq N-1 $.

$ x \; \de\ \; [1,S_k(N(m+1)+r)] \; = \;
[1,k,k+1,\dots,N,1,2,\dots,k-1,S_k(Nm+r)].$ By virtue of the Jacobi identity:
\be
 x &  = &  [k,k+1,\dots,N-1,[1,N],1,2,\dots,k-1,S_k(Nm+r)] \\
   &    &  + \; [k,k+1,\dots,N-1,N,1,1,2,\dots,k-1,S_k(Nm+r)].
\ee
Let us denote the first(second) summand in the right-hand side of the above
formula by $ a(b) $.
Due to the Jacobi identity and the defining relations:
\be
a & = & -[k,k+1,\dots,N-1,N,2,3,\dots,k-1,S_k(Nm+r)] \\
  &   & +[k,k+1,\dots,N-1,1,[1,N],2,3,\dots,k-1,S_k(Nm+r)].
\ee
Denoting the first(second) summand in the right-hand side of the above formula
by $a_1(a_2)$ and using the definition of a string we find: $ a_1 \; = \;
[k,k+1,\dots,N-1,N,S_2(Nm+r+k-2)].$ Then applying the already proven relation
${\rm d_{m+1}}$ and the automorphism $ C$ to compute the commutator $
[N,S_2(Nm+r+k-2)] $ one obtains $ a_1\; =\;0. $ For $ a_2 $ we have:
\be
a_2 & = & [k,k+1,\dots,N-1,1,1,N,S_2(Nm+r+k-2)] \\
    &   & -[k,k+1,\dots,N-1,1,N,1,2,\dots,k-1,S_k(Nm+r)]
\ee
The first term in this expression for $ a_2 $ is equal to zero because $
[N,S_2(Nm+r+k-2)]\;=\;0 $ while the second term is equal to $ -x $.

Applying the already proven relation ${\rm b_{m+1}}$ and the same reasoning as
in the computation of $a_1$
 we find that $ b\; =\; 0 $. Therefore $ x\;=\; a_1 + a_2 + b \; = \; -x $ ; $
x\; = \; 0. $

%%%%%%%%%%%%%%%%%%%%%%%%% THE FIRST ISLAND
\vspace{0.4cm}

$\underline{ Case \; {\rm e_{m+1}} }$  :  $ k+r = N+1,\; k \geq 3 ,\; r \leq
N-1 $.

Applying the Jacobi identity we obtain:
\be
x & \de\ & [1,S_k(N(m+1)+r)] \\
 & \de\ & [1,k,k+1,\dots,N,1,2,\dots,k-1,S_k(Nm+r)] \\
 &  = &  [k,k+1,\dots,N-1,[1,N],1,2,\dots,S_k(Nm+r)] \\
 &  & +[k,k+1,\dots,N,1,[1,2],3,\dots,k-1,S_k(Nm+r)] \\
 &  & + [k,k+1,\dots,N-1,N,1,2,\dots,k-1,1,S_k(Nm+r)]
\ee
Let us denote the three summands standing in the right-hand side of the last
equality in the above formula by $ a\;,b\;,c\;.$
Using the defining relations, the Jacobi identity and the definition of a
string we come to the equality:
\be
a & = & -[k,k+1,\dots,N-1,N,S_2(Nm+r+k-2)] \\
 &  & + [ k,k+1,\dots,N-1,1,1,N,2,3,\dots,S_k(Nm+r)]\; - \; x\; .
\ee
Denoting the first(second) summand in the above expression for $ a $ by $
a_1(a_2) $, and using the relation ${\rm g_m}$ we find that $ a_1 \;=\;
[k,k+1,\dots,N-1,S_2(Nm+N-2)].$ Whereas applying the definition of a string and
the relation ${\rm b_m}$ we find that $ a_2 \; = \; -a_1 $. Therefore $ a\;=\;
-x $.

For $ b $ we obtain:
\be
b & \de\ & [k,k+1,\dots,N,1,[1,2],\dots,k-1,S_k(Nm+r)] \\
  &  =  &  [k,k+1,\dots,N,1,1,2,\dots,k-1,S_k(Nm+r)]  \\
  &   & -[k,k+1,\dots,N,1,2,1,3,\dots,k-1,S_k(Nm+r)]
\ee
Denote the first(second) summand in the right-hand side of the above equality
by $ b_1(-b_2) $. Then the definition of a string and ${\rm i_m }$ enable us to
write:
\be
b_1 & = & [k,k+1,\dots,N,1,S_1(Nm+r+k-1)] \\
  & = & -2[k,k+1,\dots,N,S_1(Nm+N+1)] \\
  & = & -2S_k(N(m+1)+r+1)
\ee
Whereas ${\rm e_m} $ applied to the commutator $ [1,S_3(Nm+r+k-3)]$ entering
$b_2$ gives $ b_2\;=\; -S_k(N(m+1) +r +1). $

The relation $ {\rm e_m} $ applied to the commutator $ [1,S_k(Nm+r)] $ entering
$c$ leads to $ c \; = b_2 $.
Finally: $ x\; = \; a+b+c \; = \; -x - 2S_k(N(m+1)+r+1)\; ; \; x \; = \;
-S_k(N(m+1)+r+1). $

%%%%%%%%%%%%%%%%%%%%%%%%% THE SECOND ISLAND
\vspace{0.4cm}

$\underline{ Case \; {\rm g_{m+1}} }$  :  $ k+r = N+2,\;  r \leq N-1 $.

Applying the Jacobi identity we obtain:
\be
x & \de\ & [1,S_k(N(m+1)+r)] \\
 & \de\ & [1,k,k+1,\dots,N,1,2,\dots,k-1,S_k(Nm+r)] \\
 &  = &  [k,k+1,\dots,N-1,[1,N],1,2,\dots,S_k(Nm+r)] \\
 &  & +[k,k+1,\dots,N,1,[1,2],3,\dots,k-1,S_k(Nm+r)] \\
 &  & + [k,k+1,\dots,N-1,N,1,2,\dots,k-1,1,S_k(Nm+r)]
\ee
Let us again denote the three summands standing in the right-hand side of the
last equality in the above formula by $ a\;,b\;,c\;.$
Using the defining relations, the Jacobi identity and the definition of a
string we come to the equality:
\be
a & = & -[k,k+1,\dots,N-1,N,S_2(Nm+r+k-2)] \\
 &  & + [ k,k+1,\dots,N-1,1,1,N,S_2(Nm+k+r-2)]\; - \; x\; .
\ee
Using the relation ${\rm l_m}$ and the automorphism $C$ to compute the
commutator $ [N,S_2(Nm+r+k-2)] $ we arrive at $ a \;=\; -x $.

For $ b $ we obtain:
\be
b & \de\ & [k,k+1,\dots,N,1,[1,2],3,\dots,k-1,S_k(Nm+r)] \\
  &  =  &  [k,k+1,\dots,N,1,1,2,\dots,k-1,S_k(Nm+r)]  \\
  &   & -[k,k+1,\dots,N,1,2,1,3,\dots,k-1,S_k(Nm+r)] \\
  & = & [k,k+1,\dots,N,1,S_1(Nm+N+1)] - [k,k+1,\dots,N,1,2,1,S_3(Nm+N-1)]
\ee
Now we use the already proven relation ${\rm a_{m+1}}$ to compute the
commutator $ [1,S_1(Nm+N+1)] $ inside the first bracket above and the relation
${\rm g_m}$ - to compute the commutator $ [1,S_3(Nm+N-1)] $ inside the second
one. This gives $ b\; = \; -S_k(N(m+1)+r-1) $.

Applying ${\rm g_m}$ to the commutator $ [1,S_k(Nm+r)] $ entering $c$ we get $
c \; = \; b $.

Finally: $ x\;=\; a+b+c \; =\; -x - 2S_k(N(m+1)+r-1)\;;\;
x\;=\;-S_k(N(m+1)+r-1) $.

%%%%%%%%%%%%%%%%%%%%%%%%%%% THE FIRST CYCLIC CASE
\vspace{0.4cm}

$\underline{ Case \; {\rm i_{m+1}} }$  :  $ k=1\;, r = N $.

$x \; \de\ \; [1,S_1(N(m+1)+N] \; \de\ \; [1,1,2,S_3(N(m+1)+N-2)] \; = \;
[1,[1,2],S_3(N(m+1)+N-2)]
+ [1,2,1,S_3(N(m+1)+N-2)]$. Denoting the first(second) summand in the
right-hand side of the last formula by $a(b)$, and using the defining relations
we obtain:
\be
a & = & [2,S_3(N(m+1)+N-2)] + [[1,2],1,S_3(N(m+1)+N-2)]
\ee
Applying the already proven relation ${\rm e_{m+1} }$ to compute the commutator
$ [1,S_3(N(m+1)+N-2)] $
one gets:
\be
a & = & [2,S_3(N(m+1)+N-2)] - S_1(N(m+1) + N+1) + [2,1,S_3(N(m+1)+N-1)]
\ee
Transforming the last summand with the aid of the already proven relation ${\rm
g_{m+1} }$ we arrive at :  $ a \; = \; -S_1(N(m+1)+N+1) $.

Application of $ {\rm e_{m+1} } $ to $b$ gives $ b \; = \; a $. Hence $ x\; =
\; a+b \; = -2S_1(N(m+1)+N+1) $.

%%%%%%%%%%%%%%%%%%%%%%%%%%%%% CYCLIC GAP
\vspace{0.4cm}

$\underline{ Case \; {\rm l_{m+1}} }$ : $ 3 \leq k \leq N-1 \;, r=N $.

$ [1,S_k(N(m+1)+N)]\;=\;[k,1,S_{k+1}(N(m+1)+N-1)]\;= 0 .$ In virtue of the
already proven relation ${\rm h_{m+1} }$.

%%%%%%%%%%%%%%%%%%%%%%%%%%%%%%%% THE THIRD CYCLIC CASE
\vspace{0.4cm}

$\underline{ Case \; {\rm k_{m+1}} }$ : $ k=N\;, r=N $.
\be
x & \de\ & [1,S_N(N(m+1)+N)] \;=\; [1,N,1,S_2(N(m+1)+N-2] \\
  &  = & [[1,N],1,S_2(N(m+1)+N-2)] + [N,1,1,S_2(N(m+1)+N-2)]
\ee
Denoting the first(second) summand in the right-hand side of the last equality
by $ a(b) $ and using the defining relations we obtain:
\be
a & = & -[N,S_2(N(m+1)+N-2]+[1,1,N,S_2(N(m+1)+N-2)]\;-\;x
\ee
Transforming the second summand in the above expression for $ a $ with the aid
of the already proven relations ${\rm e_{m+1}}$  and ${\rm i_{m+1}}$ we arrive
at: $ a \;=\; -[N,S_2(N(m+1)+N-2)]+2S_2(N(m+1)+N+1) - x $.

For $b$ we have: $ b \;=\; [N,1,S_1(N(m+1)+N+1)] $. Taking into account the
already proven relation ${\rm b_{m+1}}$ one then gets: $ b \; = \;
[N,S_2(N(m+1)+N-2] $. Finally: $ x \; = \; a+b \; = -x + 2S_1(N(m+1)+N+1)\; ; x
\; = \; S_1(N(m+1) + N+ 1) $.

The remaining cases: $ {\rm c_{m+1}}\;,\; {\rm f_{m+1}}\;, \; {\rm j_{m+1}} $
are direct consequences of the definition of the strings.

Thus the induction step is proven. \begin{flushright} $ \Box $ \end{flushright}

The Lemma has obvious corollary:

%\begin{cor}
\pro Corollary  The elements $ x(m)\; \de\ \; \sum_{i=1}^N S_i(Nm) $ belong to
the centre of  \ac\ .
%\end{cor}

{\it Proof}

It follows at once from the statements $ {\rm i_m}\;,\;{\rm j_m}\;,\;{\rm k_m}$
and ${\rm l_m}$ of
the Lemma by application of the automorphism $C$, that $ [i,x(m)]\;=\;0\;,\;
1\leq i\leq N \;,\;m\geq 1 $
 \begin{flushright} $ \Box $ \end{flushright}

Now we can proceed further and turn to the proof of the proposition 1. First of
all we notice that any multiple commutator of $\e_i$-s can be converted with
the aid of the Jacobi identity into a linear combination of commutators nested
to the right, i.e. commutators of the form $ [i_1,i_2,i_3,\dots,i_{m-1},i_m] $
for some set of $1 \leq i_k \leq N$. Let us  compute the nested commutators in
the last expression starting from the innermost one: $[i_{m-1},i_m]$, and going
step by step outwards. At each step of this procedure we need to compute a
commutator of a generator with a string which is , according to the Lemma,
again a string. Therefore any commutator of the form:  $
[i_1,i_2,i_3,\dots,i_{m-1},i_m] $ is a string (may be of zero length, then it
is equal to 0). Thus any multiple commutator of the generators is a linear
combination of strings ( with integer coefficients). This finishes the proof of
the proposition 1.

In principle now we could find commutation relations among all strings using
the Jacobi identity and the result of the Lemma. Such a computation, though, is
rather cumbersome and we did it only for $N=3$. For arbitrary $N \geq 3$ in
sec.  5 we follow another route which eventually gives  a basis of the algebra
\ac\ in terms of strings and commutation relations between the elements of this
basis.

\section{Isomorphism of the algebras \ac\ and \act\ }

In this section we show that the algebra \ac\ defined in sec.2 and the algebra
\act\ defined in sec. 3 are isomorphic.

We define a linear map from \ac\ to \act\ which we call $\pi$. First, we define
this map on the generators of \ac\ as follows:
\ben \pi(\e_i) & \de\ & \St_i(1) \;\;\; \; \; 1 \leq i \leq N
\een where the vectors $\St_i(1)$ were defined in (23) in sec. 2.

Next, we define the map $\pi$ on the whole algebra \ac\ by the prescription:
\ben \pi( [a,b] ) & \de\  & [ \pi(a) , \pi(b) ]  \een

It is easy to check that the vectors $\pi(\e_i) $ satisfy the relations
(11-12). Therefore the map $ \pi : \ac\ \; \rightarrow \;  \act\ $ is a
homomorphism of Lie algebras. Now we wish to find out what is the image of the
algebra \ac\ under the action of $\pi$. Since \ac\ is spanned by strings, it is
sufficient to find images of all strings , i.e.   vectors $ \pi( S_i(r) ) \;
\;\;1\leq i \leq N $ . Using the definition of a string, and the prescription
(27), we arrive at the recursion relation:
\ben  \pi(S_i(r)) & = & [ \pi(\e_i) , \pi(S_{i+1}(r-1))  ]  \een

This recursion relation is supplemented by the initial conditions (26),
therefore we can solve it, the result being:
\ben  \pi(S_i(r)) & = & \St_i(r) \;\;\;\; 1 \leq i \leq N  \;\;,\;\;
r=1,2,\dots  \een where $ \St_i(r) $ are defined in (23-25). Since  \act\ is a
linear span of  the vectors $ \St_i(r) \;\;\;\; 1\leq i\leq N \;\;,\;\;
r=1,2,\dots $, we come to the conclusion that the image of \ac\ is the whole
algebra \act\ :  $  Im\: \pi |_{\ac\ } \; = \; \act\  $. Now we notice, that
all the vectors $ \St_i(r) \;\;\;\; (1\leq i\leq N \;\;,\;\; r=1,2,\dots) $ are
linearly independent in \act\, except that $ \sum_{i=1}^N \St_i(Nm) \; = \; 0 $
for all $ m \geq 1 $. Therefore we come to the conclusion that the kernel of
the homomorphism $\pi $ is given by:
\ben Ker\: \pi & = & linear \;\; span\;\; of\;\; \{ x(m) \}_{m\geq 1}\;\; (\de\
\; {\rm {\bf C}} \{ x(m) \}_{m\geq 1} ) \een where the elements $ x(m) \; \de\
\;  \sum_{i=1}^N S_i(Nm) $ were defined in the Corollary to the Lemma. Recall
that according to this Corollary $  {\rm {\bf C}} \{ x(m) \}_{m\geq 1} $
belongs to the centre of the Lie algebra \ac\ . Hence we conclude that \ac\ is
isomorphic to a central extension of \act\ by  $ {\rm {\bf C}} \{ x(m)
\}_{m\geq 1} $ . In order to  prove that the map  $\pi$ is an isomorphism we
have to show that $ Ker\: \pi = 0 $.

Now we have the following proposition:

\begin{prop}
For all $m\geq 1$ the central elements $ x(m)\; \de\ \; \sum_{i=1}^N S_i(Nm) $
vanish.
\end{prop}

{\it Proof}

To prove this proposition we shall first compute the commutator $ C(l,m)\; \de\
\; \\ \mbox{} [S_1(Nl),S_1(Nm)] $ for $ l,m \geq 1 $.

In virtue of the Lemma one has for $m\geq 1$:
\beqn
   [k,S_1(Nm+k-1)]\;=\; \left\{ \begin{array}{ll}
                                 -2S_1(Nm+1) & \mbox{if $k=1$} \\
                                 -S_1(Nm+k)  & \mbox{if $2\leq k\leq N-1$} \\
                                 S_N(Nm+N)   & \mbox{if $k=N$ }
                                \end{array}  \right.
\eeqn
Applying the above equality and the Jacobi identity we then get for $l,m\geq 1$
, $ 1\leq k\leq N-1 $ :
\ben
[S_k(Nl-k+1),S_1(Nm+k-1] & = &  \nonumber \\
           &    & [k,S_{k+1}(Nl-k),S_1(Nm+k-1)] \nonumber \\
           &    & +(1+\delta_{k,1})[S_{k+1}(Nl-k),S_1(Nm+k)]
\een
Using the last formula repeatedly we arrive at the equality:
\ben
C(l,m) & = & [1,S_2(Nl-1),S_1(Nm)] \nonumber \\
& &+ 2 \sum_{k=2}^{N-1} [k,S_{k+1}(Nl-k),S_1(Nm+k-1)] \nonumber \\
& &  + 2 [S_N(Nl-N+1),S_1(Nm+N-1)]
\een

The equation (31) leads besides to the following equality for $ m\geq 1 $ :
\beqn
[S_N(Nl-N+1),S_1(Nm+N-1)]\;=\;\left\{ \begin{array}{ll}
                                       S_N(N(m+1)) & \mbox{if $l=1$} \\
                                 \mbox{}      [N,S_1(Nl-N),S_1(Nm+N-1)] &-
\\
                                      -[S_1(Nl-N),S_N(Nm+N)] & \mbox{if $l\geq
2$}
                                       \end{array}  \right.
\eeqn
Combining the relations (33) and (34) we come to:
\ben
C(1,m) & = & [1,S_2(N-1),S_1(Nm)] \nonumber \\
       & & + 2 \sum_{k=2}^{N-1} [k,S_{k+1}(N-k),S_1(Nm+k-1)] \nonumber \\
       & & + 2S_N(N(m+1)) \;, \; \; \; m \geq 1   \\
C(l,m) & = & [1,S_2(N-1),S_1(Nm)] \nonumber \\
       & & + 2 \sum_{k=2}^{N} [k,S_{k+1}(Nl-k),S_1(Nm+k-1)] \nonumber \\
       & & - 2[S_1(Nl-N),S_N(Nm+N)] \;, \; \; \; m \geq 1 \;,\; l \geq 2
\een
Moreover, the Lemma and the Jacobi identity also give the following equation
for $ l,m\geq 1 $:
\ben
[S_1(Nl),S_1(Nm)]+2[S_1(Nl),S_N(Nm)]= &  &   \nonumber \\
  \mbox{}       [1,S_2(Nl-1),S_1(Nm)]+2[1,S_2(Nl-1),S_N(Nm)] &  &
\een

The next step is to compute the triple commutators of the form $
[i,S_j(p),S_k(q)] $ appearing in the equations (35-37). We shall do it as
follows. First we can compute the internal commutators $ [S_j(p),S_k(q)] $ up
to a central element using the values of commutators  $ [\St_j(p),\St_k(q)] $
in the algebra \act. Indeed, suppose we know that:
\beq
 [\St_j(p),\St_k(q)] \; = \; \sum_{l,r} F_{j,p;k,q}^{l,r} \St_l(r)
\eeq
where $ F_{j,p;k,q}^{l,r} $ are known structure constants. Then since \ac\ is a
central extension of $\tilde{\ac\ }$ (by the linear span of the set $ \{ x(m)
\}_{m\geq 1} $),  we have:

\beq
 [S_j(p),S_k(q)] \; = \; \sum_{l,r} F_{j,p;k,q}^{l,r} S_l(r)\;  + \; X
\eeq
where $ X $ is some element in the centre of \ac\ .

Next, we compute $ [i,S_j(p),S_k(q)] $ using the result of the  Lemma.

It is straightforward to find the commutators $ [\St_j(p),\St_k(q)] $ , their
computation gives the following formulas for the relevant triple commutators
for $ m \geq 1 $:
\ben
[1,S_2(N-1),S_1(Nm)] & = & 2S_1(Nm+N) \nonumber \\
\mbox{}[k,S_{k+1}(N-k),S_1(Nm+k-1)] & = & S_k(Nm)  \nonumber \\
\mbox{}[1,S_2(N-1),S_N(Nm)] & = & -S_1(Nm+N)  \nonumber \\
\mbox{}[1,S_2(Nl-1),S_1(Nm)] & = & 2S_1(N(m+l))-8(-1)^N S_1(N(m+l-2))\;\;\;
l\geq 2 \nonumber \\
\mbox{}[k,S_{k+1}(Nl-k),S_1(Nm+k-1)] & = & S_k(N(l+m))-4(-1)^N
S_k(N(m+l-2))\;\;\; l,k \geq 2  \nonumber \\
\mbox{}[1,S_2(Nl-1),S_1(Nm)] & = & -2[1,S_2(Nl-1),S_N(Nm)]
\een
Substituting these expressions into (35-37) we obtain for $ m \geq 1 $:
\ben
C(1,m) & = &  2\sum_{k=1}^N S_k(N(m+1)) \; \de\ \; 2x(m+1)\; , \nonumber \\
C(l,m) & = &  C(l-1,m+1) + 2x(l+m) - 8(-1)^N x(l+m-2)  \;\;\; ,\; l \geq 2
\een
Now taking into account that $ C(l,m)+C(m,l)\;=\;0 $ we arrive at the following
recursion relation for $ x(m) $:\beqn
mx(m)\;=\;4(m-2)(-1)^N x(m-2) \;\;\;\; m\geq 3
\eeqn
This recursion relation is supplemented by two initial conditions: $ x(2) \; =
\; 0 $ which follows from (39) and $ x(1) \; = \; 0,$ which will be shown
shortly. Solving the recursion relation (40) with these initial conditions we
get the desired result: $ x(m)\;=\;0 $.

Now we show that $ x(1) \;=\; 0 $.
\be
x(1) \; \de\ \; [1,2,\dots,N]+[2,3,\dots,1]+\dots+[N,1,\dots,N-1]
\ee
If $ N=3 $ , the equation $ x(1) = 0 $ is the Jacobi identity.
If $ N \geq 4 $ , using the defining relations of \ac\ we find:
\be
[N,1,\dots,N-1] & = & [[N,1],2,3,\dots,N-1]-[1,2,\dots,N],
\ee
\be
\mbox{} [[N,1,2,\dots,k],k+1,k+2,\dots,N-1] &= &       \\
\mbox{} [[N,1,2,\dots,k+1],k+2,\dots,N-1]-[k+1,\dots,k] & & k \leq N-3
\ee
Applying the last equation repeatedly we arrive at:
\beq
[N,1,2,\dots,N-1]\;=\;-[1,2,\dots,N]-[2,3,\dots,N]-\dots -[N-1,N,1,2,\dots,N-2]
\eeq
The Proposition is proven.
\begin{flushright}  $\Box$ \end{flushright}


In virtue of Proposition 2 $ Ker\: \pi  = 0 $ , hence the map $ \pi $ defined
in (29) is isomorphism of Lie Algebras.


\section{The Hamiltonian and the Integrals of Motion}

Now we are ready to discuss how the $sl(N)$ Onsager's algebra leads to the
existence of an infinite number of integrals of motion for a Hamiltonian which
is linear in the generators $\e_i$ of this algebra.

Suppose that we have a representation of the algebra \ac\ i.e. a set of $N$
linear operators $ \e_i \;,\; 1 \leq i \leq N $ satisfying the generalized
Dolan-Grady conditions (11,12). Consider the operator $H$ - Hamiltonian:
\ben
H & = & k_1 \e_1 + k_2 \e_2 + \dots + k_N \e_N
\een where $ k_i$ are some constants.

If we consider $ H $ as a vector in \ac\ then the  image $ \pi(H) $ of $H$
under the action of the isomorphism $ \pi $ (29) is a vector in \act\ :
\ben
\pi(H) & = & \sum_{i=1}^N k_i \St_i(1)  \\
       & = & \sum_{i=1}^{N-1}k_i( E_{ii+1} + E_{i+1i}) \; + \; k_N ( tE_{N1} +
t^{-1}E_{1N} )
\een
Now let us consider the following set of vectors in \act\ :
\ben
\tilde{I}_1 & =  & \sum_{i=1}^N k_i (\St_i(N+1) + 2\St_{i+1}(N-1)) \\
            & = & (t + (-1)^N t^{-1}) \pi(H) \\
\tilde{I}_{m} & = & \sum_{i=1}^N k_i (\St_i(Nm+1) + 2\St_{i+1}(Nm-1)) \\
            & = & ( t - (-1)^N t^{-1})^2 ( t + (-1)^N t^{-1})^{m-2} \pi(H)
\;\;\;\; m = 2,3,\dots
\een

These vectors obviously commute between themselves and with the operator $
\pi(H) $. Therefore taking the inverse of $ \pi $ we arrive to the conclusion
that the members of the following set of elements $ \{ I_m \}_{m\geq 0} $ of
the algebra \ac\ commute between themselves:
\ben
I_m & = &  \sum_{i=1}^N k_i (S_i(Nm+1) + 2S_{i+1}(Nm-1)) \; \; , \; m \geq 1 \\
 I_0 &  = &  H  \\
\mbox{} [ I_m , I_n ] & = & 0 \;\;,\; m \geq 0
\een
Thus if the conditions (11,12) are satisfied, $ H $ is a member of the infinite
family of integrals of motion in involution. Each of these integrals is a
vector in \ac\ and is expressed in terms of the operators $ \e_i $ according to
the definition of strings $ S_i(r) $ given in (13). Since the strings are
linearly independent, the vectors $ I_k $ are linearly independent in \ac\ as
well.

\section{Conclusion}

As we have seen, the original Dolan Grady relations that define Onsager's
Algebra admit a generalization. This generalization stands in the same
relationship to underlying $ sl(N) $ Loop Algebra for $ N \geq 3 $ as the
original Onsager's Algebra  -  to $ sl(2) $ Loop Algebra. The crucial property
of Dolan Grady relations - the fact that they generate an infinite series of
integrals of motion in involution - is naturally present in this
generalization.

A number of further questions present themselves. The first two of these are:
what is the analog of Onsager's Algebra for any Kac-Moody Lie Algebra and what
are the corresponding analogs of the Dolan Grady conditions? The answers are
quite straightforward and we intend to report on this in a forthcoming
publication.

 Another problem  is concerned with the representation theory of the $sl(N)$
Onsager's Algebra. The classification of finite-dimensional representations of
\ac\ should be carried out in order to obtain an analogue of the eigenvalue
formula (9). Finally we need to find examples of models with Hamiltonians of
the form (10) that satisfy the conditions (11,12) and cannot be mapped onto
free-fermions. Some special cases of the spin-chains associated with the
$sl(N)$ Chiral Potts Models \cite{sl(N)} seem to be natural candidates for such
models.

\vspace{1cm}

{\bf Acknowledgments} \\
We are very grateful to Professor B.M. McCoy for teaching us Onsagers's
Algebra,  encouragement and useful suggestions during preparation of this
paper. We are grateful to Professor J.H.H. Perk for valuable comments and to
Professors V.E.Korepin and M.Ro\v cek for their support.

\end{document}